# Utilizing the Lindemann-Gilvarry criterion for parameterization of the equations of state of noble gases Ar,Xe and Kr

Joseph Gal*
Ilse Katz Institute for Nanoscale Science and Technology,
Ben-Gurion University of the Negev, Beer Sheva ,84105 Israel

## Abstract

The solid noble gases Ar, Xe and Kr equations of state and their melting data are fitted by applying Lindemann-Gilvarry criterion. Assuming isochoric conditions in the DAC the Lindemann-Gilvarry criterion is applicable for predicting the melting curves and equations of state of the above three noble gasses. The demand that the fitting of the experimental EOS (P-V space) data will simultaneously fit the experimental cold melting data reveal a completely different bulk moduli reported in the literature. The bulk moduli $B_o$, $B_o'$ which are derived simultaneously and separately for each crystallographic phase (fcc or hcp) indicate much harder behavior contrary to very soft behavior reported previously. This explains the curvature in the melting data reported by Boehler et al. [21]. In addition, the combined approach lead to direct determination of the Grüneisen parameter $\gamma_o$, thus a safe extrapolation of the melting curve to high pressures and temperatures is achieved.

*jgal@bgu.ac.il

# 1. Introduction

In the present paper it is shown that the Lindemamm's criterion is applicable to the solid noble gases Ar, Xe and Kr. The crystal structures of the solid rare gases have been of particular interest as first principal calculations have indicated that for these van der Waals crystals the static lattice energy point cubic (fcc) and close-packed hexagonal (hcp) structures. However, calculations based on pairwise interactions have predicted greater stability for the hcp structure which is experimentally observed at pressures above 20GPa.The determination of the pressure dependent melting temperatures of solids has drawn the attention of the scientific community for many years.

In the present contribution we apply Lindemamm's criterion for prediction of melting curves, understanding that this criterion is not a theoretical model based on first principles but a phenomenological approach to the behavior of solids. We adopt and trust the Lindemann criterion improved by Gilvarry, known as Lindemann-Gilvarry (LG) criterion [1]. Prediction of the melting point at high pressures and temperatures for metals utilizing the LG criterion needs the Grüneisen parameter $\gamma$. The procedure utilizing the LG criterion together with Grüneisen parameter $\gamma$ according to the Slater model [2] often does not fit the experimental melting results. In a recent publication [3] I have proposed a different fitting procedure which takes into account simultaneously the LG criterion (P-T space) together with the equation of state (EOS P-V space). In this procedure the shock waves melting data serve as anchor to determine the actual melting curve measured in a diamond anvil cell (DAC). The LG formulation uses the bulk modulus B and its pressure derivative B' as fit parameters deduced directly from the EOS, however, the results are not unique depending on the chosen EOS. Numerous EOS are available most of them need two free parameters; the bulk moduli B and B' which are deduced from the P-V room temperature isotherm and are assigned $B_o$ and $B_o$'. Nevertheless, the fitting of the experimental data in the P-V space strongly depend on the chosen equation of state. The reported values of $B_o$ and $B_o$' span up to ~ 50% (see tables I,II) and the question remain which bulk moduli should be addressed. The bulk moduli are of basic importance for the prediction of melting curves at high pressures and temperatures for materials utilizing the LG criterion.

**Table I**: Elastic bulk modulus $B_o$ in GPa and its pressure derivative $B_o$' derived by the combined approach steps-1,2 compared to those reported in the literature.

|  | $B_o$ | $B_o$' | $B_o$ span | $B_o$'span |
|---|---|---|---|---|
| Aluminum | 73(1) | 4.45 | 72 -77  [5,10,11,13] | 4.0 – 4.54 |
| Copper | 142(2) | 4.9 | 133-142 [12,13] | 4.54 – 5.0 |
| Iron | 163(1) | 5.55 | 163-193 [14,15] | 4.2 - 5.38 |
| Uranium | 136(2) | 3.8 | 104-147 [16,17,18] | 3.8-6 |

In the case of the solid rare gasses values of $B_o$ and $B_o'$ span up to ~ 100% as demonstrated for example in the case of Krypton (Table II).

**Table II.** Equation of state elastic parameters of Krypton: volume and bulk moduli $B_o$ and its pressure derivative $B_o'$ at 300K reported in the literature. Note the span in the bulk moduli (for details see below).

| Authors | $V_o$ (Å$^3$) | $B_o$ | $B_o'$ | Pressure range (GPa) |
|---|---|---|---|---|
| Rosa et al. (2018) | 219(1) | 1.55(4) | 7.10(5) | 1-140 |
| Jephcoat at al. (1998) | 180 | 3.32 | 7.23 | 3-136 |
| Polian et al. (1089) | 240(3) | 1.4(5) | 4.3 | 2-30 |
| Errandonea et al. | 197.3 | 2.7 | 5.4 | 1-50 |
| Tian et al. calculation | 182.3 | 4.37 | 5.70 | 0-300 |

In the present contribution it is shown that the solid noble gases Ar, Xe and Kr equations of state and their melting data can be fitted by applying Lindemann-Gilvarry criterion. Different from the literature the bulk moduli derived for the fcc and hcp phases indicate less compressible materials, contrary to very soft materials reported in previous publications.

## 2. Lindemann-Gilvarry criterion

According to Lindenmann's criterion the melting temperature $T_m$ is related to the Debye temperature $\Theta_D$ as follows:

$$T_m = C \, V^{3/2} \, \Theta_D^2 \qquad (1)$$

Where V is the volume and C is a constant to be derived for each specific metal. In the Debye model the Grüneisen parameter $\gamma$ is defined by $\gamma = \partial \ln \Theta_D / \partial \ln V$. As shown by Anderson and Isaak [4] combining (1) and by inserting $V_o/V = \rho/\rho_o$, and integrating one gets the form of LG criterion of the melting temperature $T_m$:

$$T_m(\rho) = T_{mo} \exp \left\{ \int_{\rho o}^{\rho} [2\gamma - 2/3] \, d\rho/\rho \right\} \qquad (2)$$

Where $\rho_o$ is a reference density, $\rho$ is the density at the melt and $T_{mo}$ is the melting temperature at the reference density. Integrating (2) assuming that $\gamma = \gamma_o (\rho_o/\rho)^q$ and $q = 1$ one gets:

$$T_m(\rho) = T_{mo} (\rho_o/\rho)^{2/3} \exp[2\gamma_o (1 - \rho_o/\rho)] \qquad (3)$$

where $\gamma_o$ is defined as the Grüneisen parameter at ambient conditions.

Equation (3) states that if $\rho(P)$, $T_{mo}$ and $\gamma_o$ are known the melting curve $T_m(P)$ can simply be determined assuming that the relation between P and $\rho$ is known. It is well accepted that the pressure in the P-V-T space is given by:

$$P(V,T) = P_C + \gamma_{lattice} C_{v\,lattice} \rho [T-T_o+E_o/C_{v\,lattice}] + \tfrac{1}{4} \rho_o \gamma_e \beta_o (\rho/\rho_o)^{1/2} T^2 \quad (4)$$

Here $P_c$ is the cold pressure, $C_v$ is the lattice specific heat above $T_o$, $T_o$ is the ambient temperature. $C_{v,lattice}$ is taken as constant (usually at room temperature, following the approximation of Altshuler et al. [5] ), $E_o$ is the lattice thermal energy at $T_o$ and $\gamma_{lattic}$ is the lattice Grüneisen parameter. $\gamma_e$ is electronic Grüneisen parameter and $\beta_o$ is the electronic specific heat coefficient (Altshuler ref.[5] and Kormer ref.[6]). It is customary to analyze EOS and melting experiments in terms of room temperature isotherms using Murnaghan [7],Vinet [8] and Birch- Murnaghan [9] EOS for deriving $P_c$. The parameters of these equations are the ambient condition bulk modulus $B_o$ and its pressure derivative $B_o$'. In DAC experiments the volume of the tested material close to the melt must be known and is essential to apply the LG theory. Under the assumption of isochoric conditions one can directly use the measured volume in compression at room temperature. It can be also measured insitu while heating the compressed sample. In most experiments the material is compressed at room temperature and then heated to the melting point. Such experiments usually present the measured pressure ( Ruby's line shifts) at ambient temperature. These experiments ignore the thermal contribution (actual pressure) and the results are usually presented as the melting temperature vs. the cold pressure at ambient temperature, known as the cold melting curve.

The relation between $P_C$ and the density $\rho(P)$ for the room temperature isotherms frequently used are:

Murnaghan : $\quad P_c = B_o/B_o' ((\rho/\rho_o)^{-B'})$ (MUR) (5)

Vinet : $P_c = 3B_o (\rho/\rho_o)^{-2/3} [1-(\rho_o/\rho)^{1/3}] \exp\{3/2( B'-1)[1 -(\rho_o/\rho)^{1/3}]\}$ (VIN) (6)

and

Birch-Murnaghan:
$\quad P_c = 3/2\, B_o [(\rho/\rho_o)^{7/3}-(\rho/\rho_o)^{5/3}] [1+3/4(B'-4)\{ (\rho/\rho_o)^{2/3}-1\}] \quad$ BM $\quad$ (7)

Where $\rho$ is density and $B = -V(\partial P/\partial V)$ is the definition of the bulk modulus and B' is the pressure derivative of the bulk modulus ( $B'=\partial B/\partial P$). B and B' are fit parameters of the room temperature isotherm assigned as $B_o$ and $B_o'$.

It is well known that the best fit solutions are not unique and strongly depend on the chosen EOS (eq.6,7,8). This is the reason why diverse results are obtained by different authors. Thus, it make sense to introduce a different procedure in order to improve the fittings of the data in the P-V and the P-T planes using the above equations of state.

The basic assumption is that in an ideal DAC under each applied pressure, starting from ambient pressure and temperature, as raising the temperature and approaching the melt, the sample under investigation sense isochoric condition. Isochoric condition in the DAC means that the thermal pressure $P_{th}$ and the melting temperature Tm increase upon heating the sample. Thus, actual pressure eq.4 P(V,T) should be used for deriving the extrapolated melting curve.

The following four steps procedure to determine the correct melting curve (the combined approach) was proposed in Ref.3:

1. Utilizing Lindemann-Gilvarry criterion (eq.5) with $\gamma_{eff}$ as free parameter and optimizing $B_o$ and $B_o$' by choosing the appropriate EOS (out of eqs.6,7 or 8) which best fit simultaneously the experimental P-V data (isotherm 300K) and the experimental melting P-T data. Thus obtaining $P_c$ forming the cold melting curve. In LG eq.5 $Tm_o$ and $V_o$ are the melting temperature and volume at ambient pressure.

2. Adding the calculated thermal pressure $P_{th}$ to $P_c$ obtaining the LG melting curve accounting for the actual pressure (isochoric condition) sensed by the investigated sample. Demanding that the thermally corrected melting curve will include the shock wave melting data as anchors. The Grüneisen parameter $\gamma_o$ is derived accordingly.

3. Plotting the volume compression V/Vo vs. the thermally corrected melting temperatures obtained in 2.

4. Extrapolating the derived thermally corrected melting curve to high pressures and temperatures.

It has been shown that applying this procedure lead to safe extrapolation of the melting curves to high pressures and temperatures for Al,Cu, and U metals [3] and to direct determination of $B_o$, $B_o'$ as well as the Grüneisen parameters $\gamma_{eff}$ and $\gamma_o$.

In the present contribution the combined approach is applied to three compressed noble solid gases Ar,Xe and Kr. As these solids are highly compressible (soft) materials, the thermal contribution (Pth) is negligible. It is indeed a better test to start with when checking the combined approach on aluminum [3]. To conclude, the combined approach (step 1) should be applied in the case of the compressed noble gases. It is shown below that the bulk moduli $B_o$, $B_o'$ and $\gamma_{eff}$ are derived simultaneously and separately for each crystallographic phase ( fcc or hcp). Surprisingly, in the cases of Ar,Xe and Kr best fits were obtained with MUR EOS (eq.5), though VIN(eq.6) or BM (eq.7) can do the job.

## 3. Fitting Results

### Argon

The solid argon phase diagram and equation of state are depicted in Fig.1 . The melting curve of the fcc and hcp phases fitted simultaneously separately of each crystallographic phase according to the combined approach step 1. The best fits were obtained by MUR (eq.5). In the fcc region the parametrization reveal $B_o$=8.3(3)GPa and $B_o'$=3.5(1) where $\gamma_{eff}= \gamma_o=$ 1.7 (assigned 8.3/3.5/1.7). In the hcp region the melting curve reveal $B_o$=13(1) GPa and $B_o'$=3.5(1) with $\gamma_{eff}= \gamma_o=$ 2.27 (assigned 13/3.5/2.27). Equation of state P-V space shown in Fig.1b . The experimental data were reported by Ross and Mao [19] and up to 114GPa by Arrandonea et al. [20].

The melting curve of compressed Ar was measured by and reported R.Boehler et al.[21]. Depicted in Fig.1c, within the combined approach step 3, V/Vo v.s the melting temperature is calculated with the above parameters. The derived Vo is 32.6(2) Å$^3$/at. as expressed in Fig.1b .

Note that the phase transition (curvature) observed in the melting curve is also clearly pronounced in P-V space (Fig.1b).

## Xenon

The solid Xe phase diagram and equation of state are depicted in Fig.2 . The melting curve of the fcc and hcp phases fitted simultaneously separately for each crystallographic phase according to the combined approach step 1. The best fits were obtained by MUR (eq.5). In the fcc region the parametrization reveal $B_o$=7.0(2)GPa and $B_o$'=5.8(1) where $\gamma_{eff}= \gamma_o$= 2.7 assigned in Fig.2a (7/5.8/2.7). In the hcp region the melting curve reveal $B_o$=40.0(3) GPa and $B_o$'=3.0(2) with $\gamma_{eff}$= 1.0 (assigned 40/3/1). Equation of state in the P-$\rho$ space of the fcc and hcp structures are shown in Fig.2b . The experimental data were reported by Jephcoat [22] and up to 173GPa by Shindo et al. [23].

The melting of compressed Xe was measured by R.Boehler et al.[21]. Depicted in Fig.2c, within the combined approach step-3 V/Vo v.s the melting temperature is calculated with the above parameters and is depicted in the figure. $\rho_o$=4.2 g/cc (Vo=31.5Å$^3$/at.) is calculated for Xe-hcp as shown in Fig.2b (different than reported in [24,25]).

## Krypton

The solid Kr phase diagram and equations of state are depicted in Fig.3 .
The melting curve of compressed Kr was measured by R.Boehler et al.[21]. The melting curve of the fcc and hcp phases fitted simultaneously separately or each crystallographic phase according to the combined approach step 1 and are shown in Figs. 3(a),(b). The best fits were obtained with MUR (eq.5). In the fcc region the parametrization reveal $B_o$=5.5(2)GPa and $B_o$'=3.8(1) where $\gamma_{eff}= \gamma_o$= 2.8 assigned in Fig.3a (5.5/3.8/2.8). In the hcp region fitting the melting curve experimental data [21] reveal $B_o$=40(3) GPa and $B_o$'=3.2(2) with $\gamma_{eff}$ = 1.30 (assigned 40/3.2/1.30). Note that the derived density at P=0.7 GPa is 4.48 g/cc different than is given in the literature [25] (see discussion). Equation of state in P-$\rho$ space of the fcc and hcp phases are shown in Fig.3b . The experimental data relate to A. Polian [26], Jephcoat [27] and Rosa [28].

Depicted in Fig.3c, within the combined approach step 3, V/Vo v.s the temperature which is calculated with the above parameters. Vo is the value given by ref. [24] 59.1(1)Å$^3$/at. shown in Fig.3b.

**Discussion**

Already in 1964 based on low temperatures XRD measurements Lothar Meyer at al. [29] reported existence of the hcp phase in solid Ar. Computations based on pairwise interactions have predicted greater stability for the hexagonal rather than the cubic structures.

Pressure-induced fcc to hcp first order transition (martensitic) has been reported in all the solid noble gasses Ar,Xe and Kr. In this process, the low-pressure fcc phase transforms to an hcp phase as stacking disorder in the fcc lattice grows into the stable hcp domains with increasing pressure. The stacking disorder lead to defuse scattering in Xe and Kr as has been reported by Cynn et al. [30].

The experimental melting data of Ar,Xe and Kr solids were measured by R.Boehler, M. Ross, P. Soederlind and B. Boercker and reported in ref. [21]. Their results are the basis of the present contribution. The curvature in the melting curves is attributed to the phase transitions from the fcc to hcp structures based on previous EOS measurements. Indeed, the high-pressure melting data of Ar, Kr, and Xe deviate from theoretical theories [21] as the separate treatment of each crystallographic phase was not considered.

First principle theoretical calculations of the melting curve of argon by using Lindemann's criterion have been performed by Cl´oves G. Rodrigues [31]. Nevertheless these calculations miss the experimental results. Another attempt to fit the melting curve of the solid noble gases using the LG criterion to was introduced by Zheng-Hua Fang [32] reveal unacceptable bulk moduli proving that even by assuming four parameters in equation 2 ($B_o$,$B_o$',$C_1$ and q) the fcc and the hcp data can't be reasonably fitted. Both attempts report Bo<2GPa which are related to highly compressible liquids like Propylene glycol or Glycerol [34]. The reason for these discrepancies comes from ignoring the fact that the crystallographic fcc structure and the hcp phase should have been treated separately.

Pressure induced transformation from insulator to metal of solid Xe at have been subjected to many articles and is indeed related to its hcp band structure. Namely, electron transfer to the empty 5d conduction band from the full 5p valence band resulting volume reduction [33]. The similarity between the elastic properties of Xe and Kr suggest that insulator to metal transition should be expected also in compressed solid Kr.

The solid lines in Fig.1,2,3 (b) represents the fcc region as derived by the parameters obtained by best simultaneous fitting of the melting experimental data shown in Fig.1,2,3(a). All figures indicate very sharp phase transition of a first order type. The cold melting curve fitting of hcp Ar,Xe,Kr revealed relative soft materials, namely materials exhibiting $B_o <$ 50GPa (see Table III). It is assumed that in such cases the contribution of $P_{th}$ is rather small. However, adding small $P_{th}$ the actual melting curve could be somewhat flattened.

**Table III.** Summery of the derived elastic parameters of solid noble gases using the present combined approach: $V_o$, is the volume at ambient temperature and pressure. $\gamma_{eff} = \gamma_o$ is the derived Grüneisen fitting parameter (replacing $\gamma$ in eq.3)

| Solid noble Gass | Phase | $V_o$ Å$^3$/at. | $\gamma_{eff} = \gamma_o$ | $B_o$ (GPa) | $B_o'$ | $Tm_o'$ °K |
|---|---|---|---|---|---|---|
| Argon | fcc | 32.6(2) | 1.7(1) | 8.3(3) | 3.5(1) | 400 |
|  | hcp |  | 2.7(1) | 13.0(1) | 3.5(1) |  |
| Xenon | fcc | 58.15 | 2.7(1) | 7.0(2) | 5.8(1) | 400 |
|  | hcp | 31.5 | 1.0 (1) | 40(5) | 3.0(1) | 2060(10) |
| Krypton | fcc | 58.15 | 2.8(2) | 5.5(2) | 3.8(2) | 210 |
|  | hcp | 31.50 | 1.30(1) | 40(3) | 3.2(2) | 1500(20) |

**Conclusions**

By assuming isochoric conditions in the DAC the Lindemann-Gilvarry criterion is applicable for predicting the melting curves and equations of state of the three noble gasses Ar,Xe and Kr. The demand that the fitting of the experimental EOS (P-V space) data will simultaneously fit the experimental cold melting data reveal a completely different bulk moduli reported in the literature. This explains the curvature in the melting data reported by Boehler et al. [21]. In addition, the combined approach lead to determination of the Grüneisen parameter $\gamma_o$ and to a safe extrapolation of the melting curve to high pressures and temperatures in the hcp region.

**Acknowledgement**

The author gratefully acknowledge Prof. Z. Zinamon, Department of Particle Physics, Weizmann Institute of Science, Rehovot – Israel, for the many helpful illuminating discussions and comments.

Fig. 1

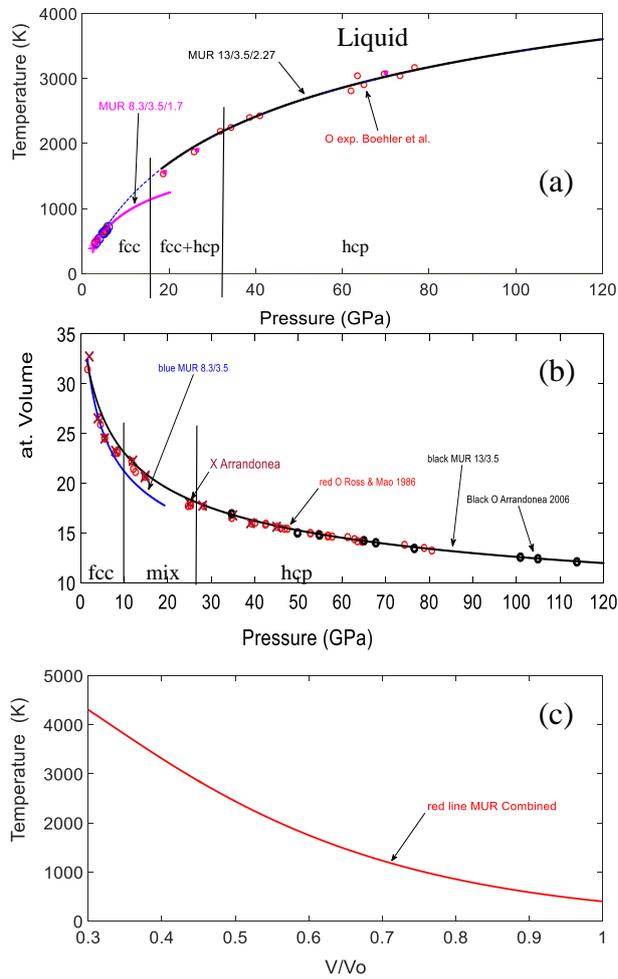

Fig. 1: **Solid argon** phase diagram and equations of state. **(a)** Melting curve of the fcc and hcp phases fitted separately of each crystallographic phase(fcc,hcp) utilizing the combined approach step 1. **(b)** Equation of state P-V space. Note that the phase transition (curvature) observed in the melting curve is also clearly pronounced in P-V space. (c) V/Vo v.s the melting temperature (for details see text). The solid lines represent fittings using Murnaghan equation of state.

Fig.2

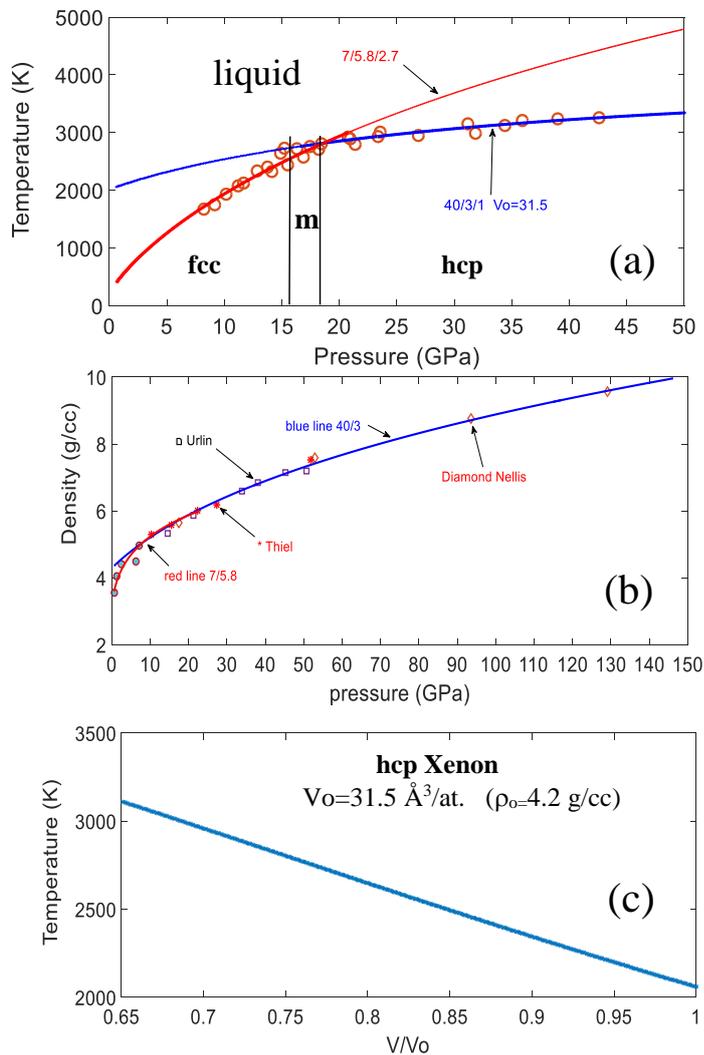

Fig.2 : **Solid xenon** phase diagram and equations of state. **(a)** Melting curve of the fcc and hcp phases fitted separately of each crystallographic structure utilizing the combined approach step 1, m indicate mixture of phases **(b)** Equation of state for the fcc and hcp phases. (c) Melting temperature v.s the relative volume V/Vo. The solid lines represent fittings using Vinet equation of state.

Fig.3

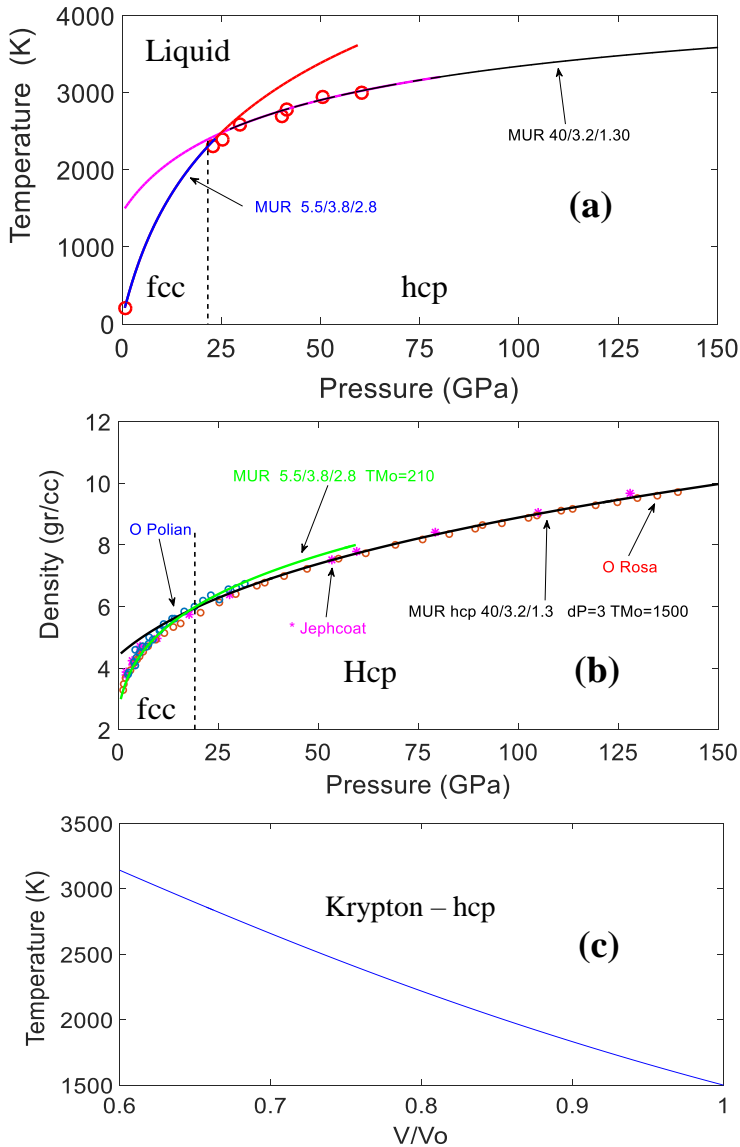

Fig. 3: **Solid Krypton** phase diagram and equations of state. **(a)** Melting curve of the fcc and hcp phases fitted separately of each crystallographic structure utilizing the combined approach step 1. **(b)** Equation of state in P-ρ plane for the fcc and hcp phases. (c) Melting temperature v.s the relative volume V/Vo. Fittings using MUR EOS.

17